# The Worldline Path Integral Approach to Feynman Graphs

Michael G. Schmidt [*]

*Institut für Theoretische Physik*
*Universität Heidelberg*
*Philosophenweg 16*
*69120 Heidelberg*

Christian Schubert [†]

*Institut für Hochenergiephysik Zeuthen*
*DESY Deutsches Elektronen-Synchrotron*
*Platanenallee 6*
*15738 Zeuthen*

### Abstract

The worldline path integral approach to the Bern-Kosower formalism is reviewed, which offers an alternative to Feynman diagram calculations in quantum field theory. Recent progress in constructing a multiloop generalization of this formalism is reported.

[*]e-mail address k22@vm.urz.uni-heidelberg.de
[†]e-mail address schubert@hades.ifh.de

# 1 Introduction

Local Quantum Field Theory (QFTH) is traditionally presented in the formalism of second quantized fields in space time $\phi_i(x^\mu)$. A set of Feynman rules allows to calculate $n$-point amplitudes in some order in the couplings inspecting a set of Feynman graphs up to a certain loop order. These rules can be conveniently derived from the Feynman path integral formalism summing over the fields $\phi_i(x^\mu)$.

String theory (STH) in contrast is usually formulated in first quantization. The basic objects in the most naive version are the space-time coordinates $x^\mu(\sigma, \tau)$ now living on a 2-dimensional worldsheet with one time and one string parameter. STH should lead to QFTH in the limit of infinite string tension $\frac{1}{\alpha'}$. This was always clear since the Veneziano model got its string interpretation, but it was worked out in detail only recently for the case of tree and one-string-loop amplitudes by Bern and Kosower [1]. They obtained a new set of rules for calculating QCD amplitudes very effectively, in particular if combined with the spinor helicity formalism, unitarity and with the use of space time supersymmetry [2]. These rules being derived from first quantized string theory look completely different from Feynman rules but indeed can be shown to be equivalent [3]. STH can be also seen as a generalized $\sigma$-model QFTH in two dimensions, in the path integral formulation one then sums over $X^\mu(\sigma, \tau)$. String loops also imply summation over different metrics and topologies of the 2-dimensional world sheet. In the limit $\frac{1}{\alpha'} \to \infty$ the $\sigma$-model QFTH loops are suppressed, and we expect a reduction to relativistic point quantum mechanics, the string loops now being reduced to sets of Feynman diagram loops. The UV divergencies of QFTH absent from STH (except for their relations to IR singularities) reappear (as moduli singularities) in the worldline limit because they are not balanced anymore by worldsheet $\sigma$-model divergencies, the physical analogue of counterterms in ordinary QFTH. A clean discussion of the $\frac{1}{\alpha'} \to \infty$ limit in the $\sigma$-model approach seems to be still lacking. It is amusing to observe that indeed in writing down heuristically the $\sigma$-model on the worldsheet the founding fathers started from the relativistic quantum mechanics for $x^\mu(\tau)$ as a $\sigma$-model on a worldline [4] and remarked that one can in principle also talk about QFTH in these terms. In practice this was rarely used [5] besides writing QM path integral expressions for propagators and fluctuation determinants.

Recently Strassler [6] proposed to do one-loop calculations using the well-known worldline (quantum mechanical) Langrangian of (spinning) particles and the bosonic and fermionic Green functions on the circle. In this way he reobtained the one-loop rules of Bern and Kosower. Since due to the superior organisation of string theory these rules are very effective in handling combinatorics and indices in particular in gauge theories, it would be highly desirable to have an extension of the rules to higher loops. However, the discussion of the $\frac{1}{\alpha'} \to \infty$ limit of $n$ loop (super)string theory [7] is complicated and has not yet led to practical results. Recently we made progress in extending the "$\sigma$-model on the worldline" approach to higher loops [8]. It already has some nice applications in QED calculations and looks very promising. After some remarks about 1-loop effective action calculations this will be the main topic of this talk.

Thus we have some hope that an alternative formulation of QFTH in terms of worldline QM will break the monopoly of the textbook approach with second quantization. Remarkably the corresponding formulation of the latter for STH – string field theory – requires the construction of a tremendous machinery [9] (and as far as perturbative calculations are concerned, one might even get along without it!). Of course the worldline approach cannot substitute string theory as a theoretical background: The choice of fields and the structure of the worldline Lagrangian needs a theoretical fundament.

|  LOCAL QFT  |  STRING THEORY  |
|---|---|
| SECOND QUANTIZED $\Phi_i(x^\mu)$ | STRING FIELD THEORY |

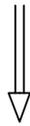

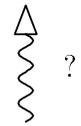

$\sum$ FEYNMAN GRAPHS      $\sum$ TOPOLOGIES METRICS

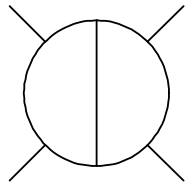            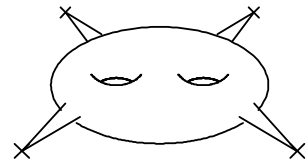

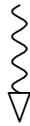            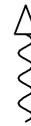

Feynman
Fradkin
Tseytlin
Polyakov

RELATIVISTIC QM ON WORLDLINE $x^\mu(\tau)$    —heuristic→    $\sigma$–MODEL $x^\mu(\sigma,\tau)$ ON WORLDSHEET
                                              ←  ?

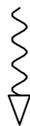 Strassler (1992, one loop)        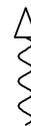

NEW SET OF RULES EQUIVALENT TO FEYNMAN RULES (TREE / ONE LOOP)   ← $\frac{1}{\alpha'} \to \infty$   FIRST QUANTIZED $x^\mu(\sigma,\tau)$

Bern, Kosower
Dunbar (1992)
Roland

## 2 One-loop worldline formalism

Let us start explaining the basic formalism inspecting one-loop effective actions written in the worldline language. A massive scalar in the loop coupled to a background gauge field $A_\mu$ and potential $V$ has the ("log det") euclidean effective action

$$\Gamma(A,V) = \frac{1}{2}\int_0^\infty \frac{dT}{T} e^{-m^2 T} \int_{x^\mu(0)=x^\mu(T)} [Dx^\mu(\tau)] \mathrm{trP}\exp\left(-\int_0^T d\tau \mathcal{L}_{WL}\right) \qquad (1)$$

with the relativistic quantum mechanical Lagrangian for the D space-time variables $x^\mu(\tau)$ depending on (Schwinger) proper time $\tau$

$$\mathcal{L}_{WL}^B = \frac{\dot{x}^{\mu 2}}{4} + igA_\mu(x(\tau))\dot{x}^\mu(\tau) + V(x(\tau)) \qquad (2)$$

In a scalar field theory $V$ would be a matrix made out of second derivatives of the field theoretical potential with respect to the field. A complex scalar just gives a factor 2 in the trace. Path integration is over all periodic functions on the circle $[0,T]$. Path ordering P applies in cases of nontrivial matrix structure of $V$ and in the case of nonabelian gauge fields $A_\mu$. This will not be considered here further since we will concentrate on electromagnetic interactions.

In the case of Dirac particles the trace gives a factor 4 and there is the usual minus sign in front. One also has to introduce a further path integration over Grassmann variables $\psi^\mu(\tau)$, related to the spin degree of freedom, adding terms to $\mathcal{L}_{WL}^B$ of eq. (2)

$$\mathcal{L}_{WL}^D = \mathcal{L}_{WL}^B + \frac{1}{2}\psi^\mu\dot{\psi}^\mu - ig\psi^\mu\psi^\nu F_{\mu\nu}(x(\tau)) \qquad (3)$$

In the case of gauge interactions $\psi^\mu$ fulfills antiperiodic boundary conditions

$$\psi^\mu(0) = -\psi^\mu(T). \qquad (4)$$

For propagators one has an open path and the scalar propagator has the form

$$<\phi(x_a)\phi(x_b)> = \int_0^\infty dT e^{-m^2 T} \int [Dx^\mu(\tau)]\exp(...)$$
$$x(0) = x_a$$
$$x(T) = x_b \qquad (5)$$

The discussion of the Dirac propagator is much more involved [10].

In calculations of the effective action one splits the coordinate path integral into "center of mass" $x_0$ and relative coordinate $y$ integration

$$\int [Dx] = \int dx_0 \int [Dy]$$
$$x^\mu(\tau) = x_0^\mu + y^\mu(\tau)$$
$$\int_0^T d\tau y^\mu(\tau) = 0 \qquad (6)$$

One then expands the external field at $x_0$ and Wick contracts as in a one-dimensional field theory on a circle (this means "stringy" approach here!). The Green functions with the proper (anti)periodic behaviour are

$$<y^\mu(\tau)y^\nu(\tau')> = -g^{\mu\nu}G_B(\tau,\tau') = -g^{\mu\nu}\left(|\tau-\tau'| - \frac{(\tau-\tau')^2}{T}\right)$$
$$<\psi^\mu(\tau)\psi^\nu(\tau')> = \frac{1}{2}g^{\mu\nu}G_F(\tau,\tau') = \frac{1}{2}g^{\mu\nu}\mathrm{sign}(\tau,\tau'). \qquad (7)$$

The free path integrals are normalized as $[4\pi T]^{-D/2}$ and 1 for the $Dy$ and $D\psi$ integration respectively. One ends up with an explicit inverse mass expansion of the 1-loop effective action. It turns out that this is a very effective way to calculate in the case of a potential [11] and also in the case of nonabelian gauge theory [12].

The Green function $G_B$ is a special case ($\rho = \frac{1}{T}$) of the Green function fulfilling [12]

$$\frac{1}{2}\frac{\partial^2}{\partial \tau^2}G(\tau,\tau') = \delta(\tau - \tau') - \rho(\tau) \tag{8}$$

with a background charge density $\rho(\tau)$ normalized as

$$\int_0^T d\tau \rho(\tau) = 1. \tag{9}$$

Our choice $\rho = \frac{1}{T}$ gives a minimal set of covariants in the effective action most directly. The usual heat kernel results are obtained with $\rho = \delta(\tau)$.

It is remarkable that the world line Lagrangian $\mathcal{L}_{WL}^D$ of eq. (3) has a globally supersymmetric form being a remnant of local SUSY after gauge fixing [13]. It is invariant under ($\eta$ Grassmann)

$$\begin{aligned}
\delta x^\mu &= -2\eta \psi^\mu \\
\delta \psi^\mu &= \eta \dot{x}^\mu.
\end{aligned} \tag{10}$$

However, SUSY is broken by the different boundary conditions for $x^\mu$ and $\psi^\mu$. Still it is convenient to introduce superfields $\hat{X}^\mu(\hat{\tau})$ on a super world-line $\hat{\tau} = (\tau, \theta)$ with constant Grassmann $\theta$

$$\hat{X}^\mu(\hat{\tau}) = x_0^\mu + \hat{Y}^\mu = x_0^\mu + y^\mu(\tau) + \sqrt{2}\theta \psi^\mu(\tau) \tag{11}$$

and a super-Lagrangian

$$\hat{\mathcal{L}}_{WL} = \frac{1}{4}D\hat{X}^\mu D^2 \hat{X}_\mu - igD\hat{X}^\mu A_\mu(\hat{X}) \tag{12}$$

where $D = \frac{\partial}{\partial \theta} - \theta \frac{\partial}{\partial \tau}$, $\int d\theta \theta = 1$. The super Green function is

$$<\hat{Y}^\mu(\hat{\tau})\hat{Y}^\nu(\hat{\tau}')> = -g^{\mu\nu}\hat{G}(\hat{\tau},\hat{\tau}') = -g^{\mu\nu}(G_B(\tau,\tau') + \theta\theta' G_F(\tau,\tau')). \tag{13}$$

From $D\hat{G}(\hat{\tau},\hat{\tau}') = -\theta \dot{G}_B(\tau,\tau') + \theta' G_F(\tau,\tau')$ we read off that unbroken SUSY ($D\hat{G} \sim \delta(\theta - \theta')$) would correspond to $\dot{G}_B = G_F$ which of course is not fulfilled with $\dot{G}_B = \text{sign}(\tau - \tau') - 2\frac{(\tau-\tau')}{T}$.

Calculating with superquantities substitutes the $\gamma$-algebra. In the component formalism this is well known for one-loop calculations [6, 14].

To examplify the above let us consider one-loop QED vacuum polarization with vanishing electron mass, and first with scalars. This can be obtained from (1) by specializing to a background $A_\mu(x) = \epsilon_{1\mu}e^{ik_1 x} + \epsilon_{2\mu}e^{ik_2 x}$ and extracting the term $\sim \epsilon_{1\mu}\epsilon_{2\nu}$. We obtain

$$\epsilon_{1\mu}\epsilon_{2\nu}(ig)^2 \int_0^\infty \frac{dT}{T}\int dx_0 \int [Dy(\tau)] \int_0^T d\tau_1 d\tau_2$$
$$\times \dot{y}^\mu(\tau_1)\exp(ik_1(x_0 + y(\tau_1)))\dot{y}^\nu(\tau_2)\exp(ik_2(x_0 + y(\tau_2)))\exp\left(-\int_0^T d\tau \frac{\dot{y}^2}{4}\right) \tag{14}$$

Contractions

$$<\dot{y}^\mu(\tau_1)\dot{y}^\nu(\tau_2)> = -g^{\mu\nu}\dot{G}'_B(\tau_1,\tau_2)$$
$$<\dot{y}^\mu(\tau_1)e^{iky(\tau_2)}> = -ik^\mu \dot{G}_B(\tau_1,\tau_2)e^{iky(\tau_2)}$$
$$<\exp(ik_1y(\tau_1))\exp(ik_2y(\tau_2))> = \exp(k_1k_2 G_B(\tau_1,\tau_2)) \qquad (15)$$

have to be performed (we use a dot (prime) for a derivative with respect to the first (second) variable). The result is

$$\epsilon_{1\mu}\epsilon_{2\nu}(ig)^2 \int_0^\infty \frac{dT}{T}(4\pi T)^{-D/2} \int_0^T d\tau_1 d\tau_2 \left(-g^{\mu\nu}\dot{G}'_B(\tau_1,\tau_2) + k_2^\mu k_1^\nu \dot{G}_B^2(\tau_1,\tau_2)\right)$$
$$\times \exp(k_1 k_2 G_{B12})\delta(k_1+k_2)(2\pi)^D. \qquad (16)$$

$\dot{G}'_B$ is partially integrated once. Rescaling $\tau_i = Tu_i$ and putting $u_2 = 0$ conveniently on the circle ($G_B = Tu_1(1-u_1), \dot{G}_B = 1-2u_1$), we end up with

$$-\frac{g^2}{(4\pi)^{D/2}}\epsilon_{1\mu}\epsilon_{2\nu}(g^{\mu\nu}k^2 - k^\mu k^\nu)(k^2)^{(D/2-2)}(\Gamma(2-D/2)/(D-1))B(D/2-1, D/2-1) \quad (17)$$

which is the well-known result.

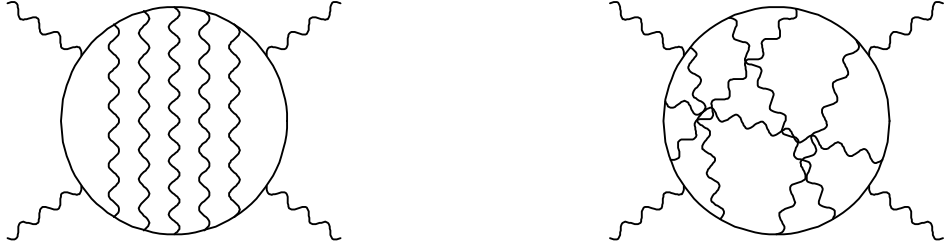

Figure 1: Feynman graphs of the same class

## 3  Multiloop worldline formalism

We now turn to our main goal, the construction of worldline Green functions for multiloop Feynman diagrams [8]. In particular we will discuss such Green functions for the class of graphs of fig. 1 where both the inner and outer vertices can be shifted arbitrarily. There will be one common $x^\mu(\tau)$ – Green function for each class like in string theory independent of the number and position of outer vertices and also independent of the couplings ($\phi^3$, Yukawa, QED...). It is not obvious that such Green functions exist:

- The Feynman graph (except in the 1-loop case) is not a one-dimensional manifold because the vertices constitute singularities (String theory regularizes these singularities, but at the price of introducing an additional dimension, and also changing the regular part of the manifold. The procedure is thus very different from the usual

"blowing up" of singularities well-known from orbifolds. However, one may hope that those singularities are irrelevant for many practical purposes, as is the case in orbifold theory).

- The parametrization along the worldlines is not unique.

We will demonstrate that such Green functions exist. For simplicity, we first consider massive $\phi^3$ theory at the two-loop level. One simply has to insert a free propagator into the circle integral over $x(\tau)$ whose endpoints $x_a = x(\tau_a)$ and $x_b = x(\tau_b)$ are on the circle and one has to integrate over $\tau_a, \tau_b, \bar{T}$ (fig. 2).

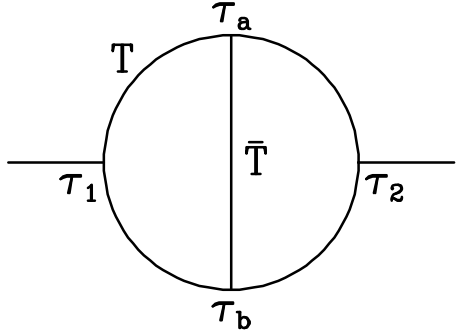

Figure 2: Two loop graph

Neglecting background couplings for a while we have

$$\int_0^\infty \frac{dT}{T} \int_0^\infty d\bar{T} e^{-m^2(T+\bar{T})} \int_0^T d\tau_a \int_0^T d\tau_b \int_{x(0)=x(T)} [Dx(\tau)] \quad (18)$$
$$\times \exp\left[-\int_0^T d\tau \frac{\dot{x}^2}{4}\right] \int_{\substack{\bar{x}(0)=x(\tau_a) \\ \bar{x}(\bar{T})=x(\tau_b)}} [D\bar{x}(\bar{\tau})] \exp\left[-\int_0^{\bar{T}} d\bar{\tau} \frac{\dot{\bar{x}}^2}{4}\right].$$

The free propagator path integral over $\bar{x}(\bar{\tau}) = x_a + (x_b - x_a)\frac{\bar{\tau}}{\bar{T}} + y(\bar{\tau})$ with $y(\bar{T}) = y(0) = 0$ of course just gives

$$\int_0^\infty d\bar{T}(4\pi\bar{T})^{-D/2} \exp(-(x(\tau_a) - x(\tau_b))^2/4\bar{T})e^{-m^2\bar{T}}, \quad (19)$$

the well-known representation of the QFTH boson propagator. This could have been also obtained by first writing down the one-loop integral with two background $\phi$ insertions (the potential $V$ of eq. (2)) and expressing the QFTH contraction $<\phi(x_a)\phi(x_b)>$ in the form (19). The exponent in (19) is quadratic in $x(\tau)$ like $\frac{\dot{x}^2}{4}$ in the free Lagrangian,

$$(x_a - x_b)^2 = \int_0^T d\tau d\tau' x(\tau)(\delta(\tau-\tau_a)-\delta(\tau-\tau_b))(\delta(\tau_a-\tau')-\delta(\tau_b-\tau'))x(\tau') =: xP_{ab}x \quad (20)$$

and can be seen as part of a Gaussian integral. Thus we obtain a new Green function $G_B^{(1)}$ on the circle defined by

$$\int_0^T d\tau' d\tau'' G_B^{(1)}(\tau,\tau') \left(\delta(\tau'-\tau'')\partial_{\tau''}^2 - \frac{P_{ab}(\tau',\tau'')}{\bar{T}}\right) y(\tau'') = y(\tau) \quad (21)$$

This can be easily solved as

$$G_B^{(1)}(\tau,\tau') = G_B(\tau,\tau') + \frac{1}{2}\frac{[G_B(\tau,\tau_a) - G_B(\tau,\tau_b)][G_B(\tau',\tau_a) - G_B(\tau',\tau_b)]}{\bar{T} + G_B(\tau_a,\tau_b)}. \tag{22}$$

The Gaussian integral normalization (below eq. (7)) changes to

$$(4\pi T)^{-D/2}(1 + \frac{1}{\bar{T}}G_B(\tau_a,\tau_b))^{-D/2}. \tag{23}$$

With $n$ outer vertices with momenta $p_i$ on the circle line (i.e. $e^{ip_i x(\tau_i)}$ vertex operator insertions) we derive with the (coherent state) contraction rules (15)

$$\int_0^\infty \frac{dT}{T}\int_0^\infty d\bar{T}e^{-m^2(T+\bar{T})}[4\pi]^{-D}\int_0^T d\tau_a \int_0^T d\tau_b [T\bar{T} + TG_B(\tau_a,\tau_b)]^{-\frac{D}{2}}$$

$$\times \prod_{i=1}^n \int_0^T d\tau_i \exp\left[\sum_{k<l} G_B^{(1)}(\tau_k,\tau_l)p_k p_l + \frac{1}{2}\sum_{k=1}^n G_B^{(1)}(\tau_k,\tau_k)p_k^2\right] \tag{24}$$

where the $G_B^{(1)}(\tau_k,\tau_k)$ are due to nonvanishing selfcontractions.

This indeed sums all two-loop diagrams with $n$ legs on the outer loop. One can check that this agrees with Feynman rule expressions.

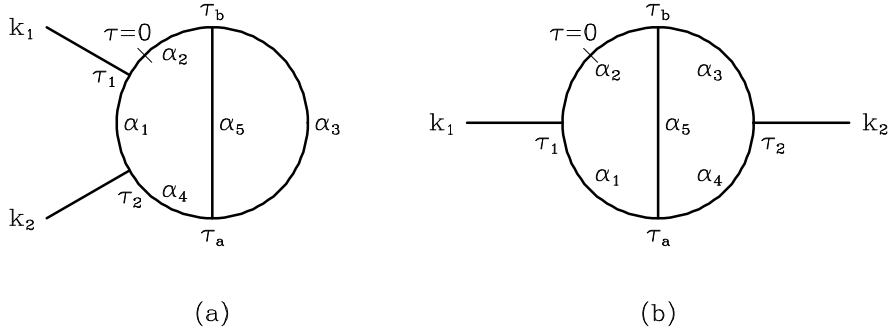

Figure 3: The two different orderings of the two loop graph

E.g. for two loops the expressions corresponding to the two diagrams fig. 3a) and 3b) in Feynman parametrization derived from the well-known network resistance analog model have the Koba-Nielsen-type form [15]

$$\int_0^\infty d\hat{T}[4\pi]^{-D}e^{-m^2\hat{T}}\prod_{i=1}^5 \int d\alpha_i \delta(\hat{T} - \sum_{i=1}^5 \alpha_i)\left[P^{(a)}(\alpha_i)\right]^{-\frac{D}{2}}\exp\left[-Q^{(a)}(\alpha_i)k^2\right], \tag{25}$$

with

$$P^{(a)} = \alpha_5(\alpha_1 + \alpha_2 + \alpha_3 + \alpha_4), \qquad P^{(a)}Q^{(a)} = \alpha_1[\alpha_5(\alpha_2 + \alpha_3 + \alpha_4) + \alpha_2\alpha_3 + \alpha_3\alpha_4], \tag{26}$$

and

$$P^{(b)} = \alpha_5(\alpha_1 + \alpha_2 + \alpha_3 + \alpha_4) + (\alpha_1 + \alpha_2)(\alpha_3 + \alpha_4),$$
$$P^{(b)}Q^{(b)} = \alpha_5(\alpha_2 + \alpha_3)(\alpha_1 + \alpha_4) + \alpha_1\alpha_2(\alpha_3 + \alpha_4) + \alpha_3\alpha_4(\alpha_1 + \alpha_2).$$

They look rather different from each other. However, after transforming from Feynman parameters to Schwinger parameters

$$\alpha_1 = \tau_1 - \tau_2, \ \alpha_2 = T - \tau_1 + \tau_b, \ \alpha_3 = \tau_a - \tau_b, \ \alpha_4 = \tau_2 - \tau_a, \ \alpha_5 = \bar{T} \quad \text{for a)} \quad (27)$$

and

$$\alpha_1 = \tau_1 - \tau_a, \ \alpha_2 = \tau_b + T - \tau_1, \ \alpha_3 = \tau_2 - \tau_b, \ \alpha_4 = \tau_a - \tau_2, \ \alpha_5 = \bar{T} \quad \text{for b)} \quad (28)$$

and rewriting the result in terms of the one-loop Green function $G_B$, they indeed assume the same form, and one can identify [8]

$$P = T\bar{T}[1 + \frac{1}{T}G_B(\tau_a, \tau_b)], \quad Q = G_B^{(1)}(\tau_1, \tau_2) - \frac{1}{2}G_B^{(1)}(\tau_1, \tau_1) - \frac{1}{2}G_B^{(1)}(\tau_2, \tau_2) \quad . \quad (29)$$

In our approach, the both cases therefore correspond merely to different regions of the $\tau_1, \tau_2$ – integral. The worldline Green function approach can be easily generalized to $m$ inner propagator insertions and leads to Green functions

$$G_B^{(m)} = G_B(\tau_1, \tau_2) + \frac{1}{2}\sum_{k,l=1}^{m}[G_B(\tau_1, \tau_{a_k}) - G_B(\tau_1, \tau_{b_k})]A_{kl}^{(m)}[G_B(\tau_{a_l}, \tau_2) - G_B(\tau_{b_l}, \tau_2)] \quad (30)$$

with the $m \times m$ matrix $A^{(m)}$ defined by

$$A^{(m)^{-1}} = \bar{T} - \frac{1}{2}B, \ \bar{T}_{kl} = \bar{T}_k\delta_{kl}, \ B_{kl} = G_B(\tau_{a_k}, \tau_{a_l}) - G_B(\tau_{a_k}, \tau_{b_l}) - G_B(\tau_{b_k}, \tau_{a_l}) + G_B(\tau_{b_k}, \tau_{b_l})$$
$$(31)$$

and with a path integral normalization factor $N^{(m)\frac{D}{2}}$, where

$$N^{(m)} = \text{Det}(A^{(m)}). \quad (32)$$

This is appropriate for the case of the QED $\beta$-function considered in the following. Note that there is just one Green function comprising all $m$ propagator configurations. Inner and outer vertices can be freely shifted along the circle line. The Green function on an inner line or with one argument on the circle and one on the inner line is needed if we also allow for background insertion on the inner line as is natural in $\phi^3$ theory. Just introducing a background term in the $D\bar{x}$-integral in (18) would lead to an asymmetric treatment. However, in the two-loop case of fig. (2) one better extends the expression (24) into an expression symmetric in all three lines [8].

## 4 QED $\beta$-function

In the case of QED the photons have a coupling

$$\int d\hat{\tau}(ieD\hat{X}^\mu(\hat{\tau})A_\mu(\hat{X}(\hat{\tau}))) = -ie\int_0^T d\tau(\dot{x}^\mu(\tau)A_\mu(x(\tau)) - \psi^\mu(\tau)\psi^\nu(\tau)F_{\mu\nu}(x(\tau))) \quad (33)$$

where $\hat{X}$ is a superconfiguration on the superworldline $(\tau, \theta)$. The photon propagator connecting two points on the circle (the closed electron line), fig. 2, in ordinary QFTH is a contraction of two $A_\mu(\hat{X}(\hat{\tau}))$, and hence is also written in superfields $\hat{X}(\hat{\tau})$. Altogether in the Feynman gauge - most appropriate if one wants to extend the scalar results - we have a photon insertion

$$-\frac{g^2}{2}\frac{\Gamma(\lambda)}{4\pi^{\lambda+1}}\int_0^T d\tau_a \int d\theta_a \int_0^T d\tau_b \int d\theta_b \frac{D\hat{X}_a^\mu D\hat{X}_{b\mu}}{((\hat{X}_a - \hat{X}_b)^2)^\lambda} \qquad (34)$$

with $\lambda = D/2 - 1$.

The scalar QED version ($\hat{X} \to x, D\hat{X} \to \dot{x}$) of expression (34) is well known from first-order corrections to scalar Wilson loops [16]. The denominator in (34) can be written in exponential form as in eq. (19) and we again have a Gaussian expression, now in supervariables. The algebra is identical to the scalar case and the modified superpropagator $\hat{G}^{(1)}(\hat{\tau},\hat{\tau}')$ is obtained by just substituting $G_B$ by $\hat{G}$ everywhere; this also holds for the determinant factor. The outer vertices obtained from the background coupling (33) of course also contain the super $\hat{X}$. In the case of effective action calculations and also for the purpose of calculating the QED $\beta$-function coordinate gauge (Fock-Schwinger gauge) for the external gauge field is very convenient [11]

$$A_\mu(x_0 + \hat{Y}) = \int_0^1 d\eta \eta \hat{Y}^\rho F_{\rho\mu}(x_0 + \eta \hat{Y}) \qquad (35)$$

where $F_{\rho\mu}$ can be covariantly Taylor expanded. We only need the constant $F_{\mu\nu}$ term $\frac{1}{2}\hat{Y}^\rho F_{\rho\mu}(x_0)$ in (35).

For the one-loop case already considered before the $F_{\mu\nu}F^{\mu\nu}$ term for $D = 4$ is obtained as

$$\begin{aligned}\Gamma_{F^2}^{(1)} &= -2\int_0^\infty \frac{dT}{T}e^{-m^2 T}\int dx_0 \int [D\hat{Y}]\exp(-\int d\hat{\tau}\hat{\mathcal{L}}_0) \\ &\quad \times \left(\frac{-g^2}{2}\right)\int d\hat{\tau}_1 d\hat{\tau}_2 \left[-\frac{1}{4}D\hat{Y}_1^\mu \hat{Y}_1^\nu F_{\mu\nu} D\hat{Y}_2^\alpha \hat{Y}_2^\beta F_{\alpha\beta}\right]\end{aligned} \qquad (36)$$

which after contractions and partial integrations leads to ($\dot{\hat{G}} = D\hat{G}$)

$$\frac{g^2}{2}\int_0^\infty \frac{dT}{T}(4\pi T)^{-2}e^{-m^2 T}\int d\hat{\tau}_1 d\hat{\tau}_2 \dot{\hat{G}}_{12}\dot{\hat{G}}_{21}\int dx_0 F_{\mu\nu}F^{\mu\nu}. \qquad (37)$$

The $d\hat{\tau}_1 d\hat{\tau}_2$ integration we can easily do, it just gives $-\frac{2}{3}T^2$. Pauli-Villars regularization of (37)

$$\int_0^T \frac{dT}{T}e^{-m^2 T} \sim \ln\frac{\Lambda^2}{m^2} \qquad (38)$$

allows to read off

$$\begin{aligned}(Z_3 - 1)^{(1)} &= \frac{2}{3}\frac{\alpha}{\pi}\left(-\frac{1}{2}\ln\frac{\Lambda^2}{m^2}\right) \\ \beta^{(1)}(\alpha) &= \frac{2}{3}\frac{\alpha^2}{\pi} \quad (\alpha = \frac{e^2}{4\pi})\end{aligned} \qquad (39)$$

In the case of two loops (fig. 2) we have similarly

$$\begin{aligned}\Gamma_{F^2}^{(2)} &= -2(4\pi)^{-4}\int dx_0 \int_0^\infty \frac{dT}{T}e^{-m^2 T}T^{-2}\int_0^\infty d\bar{T}\int d\hat{\tau}_a d\hat{\tau}_b[\bar{T} + \hat{G}_{Bab}]^{-2} \\ &\quad \times (-\frac{g^2}{2})\frac{g^2}{2}\int d\hat{\tau}_1 d\hat{\tau}_2 \frac{1}{4}\langle D\hat{Y}_1^\mu F_{\mu\nu}Y_1^\nu D\hat{Y}_2^\alpha F_{\alpha\beta}\hat{Y}_2^\beta D\hat{Y}_a^\lambda D\hat{Y}_{b\lambda}\rangle\end{aligned} \qquad (40)$$

Two types of contractions appear in the bracket, using $1 \leftrightarrow 2$ and $a \leftrightarrow b$ symmetry:

$$\left(4 \cdot 2 \hat{G}_{ab}^{(1)\prime} \hat{G}_{12}^{(1)\cdot} \hat{G}_{21}^{(1)\cdot} + 8 \hat{G}_{b2}^{(1)} \hat{G}_{21}^{(1)\cdot} \hat{G}_{1a}^{(1)\prime}\right) F_{\mu\nu} F^{\mu\nu} \tag{41}$$

These are super-Green functions obtained from (22) and the integrations in eq. (40) are over supervariables. If one does not like to keep track of the proper ordering of $\theta$-variables, one can also proceed in the following equivalent way:

(i) Do the calculation for scalars instead of Dirac fields;

(ii) eliminate $G_B^{(1)\prime}$ by partial integration;

(iii) substitute according to the cycle rule for chains of $\dot{G}_B$'s with a closed cycle of indices:

$$(\text{Closed Cycle of } \dot{G}_B) \Longrightarrow (\text{Closed Cycle of } \dot{G}_B) - (\text{Closed Cycle of } G_F)$$

We then first perform the $\tau_1, \tau_2$ integrations using identities such as

$$\int_0^1 du_2 \dot{G}_{B_{12}} \dot{G}_{B_{23}} = 2 G_{B_{13}} - \frac{1}{3} \qquad (\tau = Tu)$$
$$\int_0^1 du_2 G_{B_{12}} G_{B_{23}} = -\frac{1}{6} G_{B_{13}}^2 + \frac{1}{30} \tag{42}$$

and similar ones for the $G_F$. Those all follow from the following master identities [17], which express the kernel of the integral operator $\left(\frac{\partial}{\partial \tau}\right)^{-n}$ – with the boundary conditions in question – in terms of the Bernoulli and Euler polynomials $B_n, E_n$:

$$\int_0^1 du_2 ... du_n \dot{G}_{B_{12}} \dot{G}_{B_{23}} ... \dot{G}_{B n(n+1)} = -\frac{2^n}{n!} B_n(|u_1 - u_{n+1}|) \text{sign}^n(u_1 - u_{n+1}) \tag{43}$$
$$\int_0^1 du_2 ... du_n G_{F_{12}} G_{F_{23}} ... G_{F n(n+1)} = \frac{2^{n-1}}{(n-1)!} E_{n-1}(|u_1 - u_{n+1}|) \text{sign}^n(u_1 - u_{n+1})$$

The right hand sides of those master identities can be reexpressed in terms of $G_B, \dot{G}_B$, and $G_F$.

Substituting $\beta = (1 + \frac{\bar{T}}{G_{ab}})^{-1}$ for $\bar{T}$ the $\beta$ dependence turns out to be second-order polynomial and can be easily integrated out. We are left with the $\tau_a, \tau_b$ integrations but those are trivial: The various terms are proportional to $G_{B_{ab}}^{-n}$, $n = 0, 1, 2$, but adding them up the singular terms with $n = 1, 2$ cancel, and we are left with a constant piece!
The final result is

$$\Gamma^{(2)} = -(4\pi)^{-4} g^4 \int_0^\infty \frac{dT}{T} e^{-m^2 T} \int dx_0 F_{\mu\nu} F^{\mu\nu} \tag{44}$$

which is Pauli-Villars-regularized to

$$\Gamma_{reg}^{(2)} = -\ln\frac{\Lambda^2}{m^2}(4\pi)^{-4} g^4 \int dx_0 F_{\mu\nu} F^{\mu\nu} \tag{45}$$

and gives a two-loop contribution to the $\beta$-function

$$\beta^{(2)}(\alpha) = \frac{\alpha^3}{2\pi^2} \tag{46}$$

in agreement with the known result.

# 5  Concluding remarks

We have given an introduction to a new and effective method of doing calculations in QFT. Of course the formalism presented here still needs to be shaped by performing more demanding calculations, but one can already see several advantages in comparison with traditional Feynman graph calculations:

(i) The formalism sums up whole classes of Feynman diagrams by letting inner and outer vertices move along lines. This will be increasingly important at higher loop orders.

(ii) Thus (as in our last example) singularity cancellations are easier to see.

(iii) Worldline supersymmetry allows to treat the Dirac particles in a loop almost like scalar particles. There is no $\gamma$–algebra to perform.

(iv) Calculations should be easy to computerize.

(v) In certain cases like in the QED-$\beta$-function calculation and in effective-action calculations all integrations are polynomial. There are no momentum integrations.

Presently we have several projects proceeding based on the new method:
- the three- (and higher)-loop calculation of the $\beta$-function in QED;
- effective action calculations at the two-loop level;
- a systematic translation of Feynman vertices into worldline language.
We hope to report on this in the near future.